\documentclass[aps,pre,floats,preprint,showpacs,superscriptaddress]{revtex4}

\usepackage{graphicx,epsfig}
\usepackage{rotate}

\usepackage{times}
\usepackage{graphics,dcolumn,bm,fleqn,epic,eepic,float}
\usepackage{amssymb,amsmath,multirow,rotate,color}

\newcommand{\be}{\begin{equation}}
\newcommand{\ee}{\end{equation}}

\begin{document}

\title{Synchronization in Random Geometric Graphs}

\author{Albert D\'{\i}az-Guilera}

\affiliation{Departament de F\'{\i}sica Fonamental,
University of Barcelona, Barcelona E-08028, Spain}

\affiliation{Institute for Biocomputation and Physics of Complex
Systems (BIFI), University of Zaragoza, Zaragoza E-50009, Spain}

\author{Jes\'{u}s G\'{o}mez-Garde\~{n}es}

\affiliation{Scuola Superiore di Catania,
University of Catania, Catania I-95123, Italy}

\affiliation{Institute for Biocomputation and Physics of Complex
Systems (BIFI), University of Zaragoza, Zaragoza E-50009, Spain}

\author{Yamir Moreno}\thanks{To whom correspondence should be addressed. E-mail: yamir@unizar.es}

\affiliation{Institute for Biocomputation and Physics of Complex
Systems (BIFI), University of Zaragoza, Zaragoza E-50009, Spain}

\affiliation{Department of Theoretical Physics, University of Zaragoza, Zaragoza E-50009, Spain}

\author{Maziar Nekovee}

\affiliation{Complexity Group, BT Research and Centre for Computational Science, University College London, United Kingdom}

\date{\today}

\begin{abstract}

In this paper we study the synchronization properties of random geometric graphs. We show that the onset of synchronization takes place roughly at the same value of the order parameter that a random graph with the same size and average connectivity. However, the dependence of the order parameter with the coupling strength indicates that the fully synchronized state is more easily attained in random graphs. We next focus on the complete synchronized state and show that this state is less stable for random geometric graphs than for other kinds of complex networks. Finally, a rewiring mechanism is proposed as a way to improve the stability of the fully synchronized state as well as to lower the value of the coupling strength at which it is achieved. Our work has important implications for the synchronization of wireless networks, and should provide valuable insights for the development and deployment of more efficient and robust distributed synchronization protocols for these systems.

\end{abstract}


\maketitle

\section{Introduction}

Synchronization phenomena constitute one of the most striking examples of the emergent collective behavior that arise in many fields of science, ranging from natural to social and artificial systems 
[Winfree, 1990; Strogatz, 2003;Manrubia et al., 2004].
They have been intensively studied during the last several decades not only from an academic point of view, but also due to their applications in many man-made systems. An emerging line of research in which synchronization processes plays a key role is that of wireless communication networks. Synchronization processes in wireless systems naturally arise when routing and information flow algorithms establish a universal coordinated time, thus requiring the synchronization of the clocks of the nodes of the wireless network.  Additionally, due to the finiteness of communication channels, the access times of different users should be desynchronized when their number is large, which on its turn, is a function of the number of wireless devices accessing the available resources, i.e, of the topology of the underlying graph.

The description of the topological properties of wireless ad-hoc systems is not easy as they, unlike wired networks, are created on the fly to perform a task, such as information routing, environmental sensing, etc 
[Hekmat, 2006]
Furthermore, the topology of these networks can be changed dynamically to achieve a desired functionality. With the rapid growth of the number of portable devices and the increased popularity of wireless communication, it is expected that these types of networks will play a key role as a building block of the next generation Internet. On the other hand, from the perspective of fundamental research, these systems provide a clear-cut example of highly dynamic, self-organizing complex networks. It is therefore natural to ask how networks that topologically resemble the features of wireless settings compare with other architectures as far as synchronization processes on top of them concern.

The entangled structural and dynamical complexity of synchronization phenomena in wireless networks makes the problem difficult to tackle in fine details. Approximations that shed light on the general dynamical behavior of similar systems are thus called for. In this paper, we study the synchronization of Kuramoto oscillators on random geometric graphs (RGGs), the latter being a plausible representation of the architecture of the system when the time scale associated to the dynamics is much faster than that associated to changes in the underlying topology. We address the problem by considering the limiting situations of the onset of synchronization and the stability of the fully synchronized state. Our results show that Kuramoto oscillators achieve complete synchronization at very large coupling values as compared to their random graph counterparts. Moreover, we also show that the fully synchronized state is less stable in RGGs than in random and scale-free networks with the same number of nodes and average connectivity. Finally, we propose a mechanism by which the synchronization properties of RGGs can be greatly improved at low costs.

\section{Network model and Dynamics}

In what follows, we are interested in exploring the dynamical behavior at both the onset and the fully synchronized behavior. The first part of the phase diagram can only be explored by invoking a specific dynamics, while the stability of the fully synchronized state can be generically studied for a wide class of dynamical systems. The latter is achieved by exploring the spectral properties of the Laplacian matrix that completely describe the topology of the system under study. Using recent results of the Master Stability Function formalism 
[Barahona \& Pecora, 2002; Nishikawa et al., 2003; Motter et al., 2005; Donetti et al., 2005; Zhou et al., 2006;
Chavez et al., 2005],
one is able to reduce the problem of inspecting the stability of the completely synchronized state to an eigenvalue problem where only the topological properties of the substrate graph are relevant.

Let us then start by exploring the onset of synchronization. To this end, we consider that the  $N$ nodes of the RGG are oscillators whose dynamics are described by the Kuramoto model (KM). The KM 
[Kuramoto, 1984; Acebr\'{o}n et al., 2005]
is a model of phase oscillators
coupled through the sine of their phase differences. This
model owes most of its success to the plenty of analytical insights
that one can get through the mean-field approximation originally
proposed by Kuramoto in which nodes are globally coupled, i.e., according to an all-to-all topology. The model has also been recently used to study the synchronization of phase oscillators in complex networks 
[Moreno \& Pacheco, 2004; Oh et al., 2005; McGraw \& Menzinger, 2005; Arenas et al., 2006; G\'{o}mez-Garde\~{n}es et al., 2007a; G\'{o}mez-Garde\~{n}es et al., 2007b].
In the KM formalism, the phase of the $i$-th unit, denoted by $\theta_i(t)$, evolves in time according to
\begin{equation}
\frac{d\theta_i}{dt}=\omega_i + \lambda \sum_{j} A_{ij}\sin(\theta_j-\theta_i) \hspace{0.5cm} i=1,...,N
 \label{ks1}
\end{equation}
\noindent where $\omega_i$ stands for its natural frequency,
$\lambda$ is the coupling strength between
units and $A_{ij}$ is the connectivity matrix ($A_{ij}=1$ if $i$ is
linked to $j$ and $0$ otherwise). The model can be solved in terms of
an order parameter $r$ that measures the extent of synchronization in
a system of $N$ oscillators as:
\begin{equation}
re^{i\Psi}=\frac{1}{N}\sum_{j=1}^{N} e^{i\theta_j}
 \label{r_kura}
\end{equation}
\noindent where $\Psi$ represents the average phase of the system. The
parameter $r$ takes values $0\le r \le 1$, being $r=0$ the value of
the incoherent solution and $r=1$ the value for total synchronization.

The topological properties of the wireless system contained in the matrix $A_{ij}$ is modeled by considering that the connections among the nodes of the network vary at a time scale much slower than the time scale associated to the dynamics. Therefore, one can consider the underlying graph as static. In order to generate the network, we consider a set of nodes distributed in a two-dimensional plane. To each of these nodes, a maximum transmission range is assigned, such that a node is linked  to only those nodes within a circle of radius equal to the maximum transmission range 
[Nekovee, 2007].
Repeating this procedure for all nodes in the network, the topology of the resulting network can be described as a two-dimensional random geometric graph 
[Dall \& Christensen, 2002; Penrose, 2003].
Random geometric graphs have been used before in the study of continuum percolation and have been revitalized recently in the context of wireless ad hoc networks 
[Glauche et al., 2004; Krause et al., 2006].
They have a distribution of connectivity, $P(k)$, that like
Erd\"os-R\'eny random graphs (ER) 
[Dorogovtsev \& Mendes, 2003; Bornholdt \& Schuster, 2003; Boccaletti et al., 2006],
is peaked at an average value $\langle k \rangle$ with a finite variance.

On the other hand, other topological characteristics of the RGGs
are very far from typical random networks, namely their clustering coefficent $\langle C\rangle$ and
the average path length $\langle L\rangle$. The clustering coefficient is defined as the probability that
two neighbors of a given node are also connected betweem them, whereas the
average path lenght accounts for the average distance between every pair of nodes in the
network, being the distance between two neighbors set to $1$. As introduced before, a RGG
is constructed in a physical space and hence when two nodes are physically close to a
third one (and hence both share a link with it), the probability that they are also
physically close is high. Therefore the value of the the clustering coefficent is much higher than the typical values found in random ER
networks with the same number of nodes, $N$, and links, $N_l$. Furthermore,
the absence of links between physically separated nodes prevents the existence of
shortcuts in the network and hence RGGs have a much larger average path length than the ER graphs with the same $N$ and $N_l$.

\begin{figure}[t]
\begin{center}
\epsfig{file=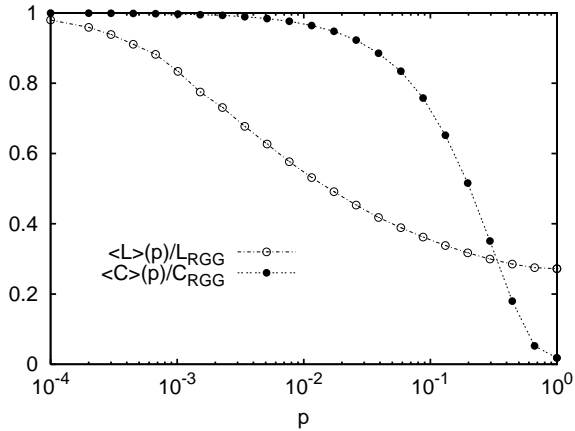,width=2.3in,angle=-90}
\end{center}
\caption{Average clustering coefficent $\langle C\rangle$ and average path
    length $\langle L\rangle$ of the rewired RGG network as a function of the
    fraction of rewired links $p$. Both quantities are normalized to the
    corresponding values of the original RGG network ($C_{\mbox{\scriptsize{RGG}}}=0.61$ and
    $L_{\mbox{\scriptsize{RGG}}}=11.55$).}
\label{fig1}
\end{figure}

To unravel how the above topological features of RGGs affect the synchronization of phase
oscillators we have constructed a family of networks whose structural characteristics
are varied from those of  RGG to those found in random graphs such as ER networks.
For this purpose a rewiring process is performed on the RGGs. The rewiring is made as follows: we consider one pair of connected nodes ($i$,$j$) and with
probability $0<p<1$ we remove the link and add a new one from one of the nodes $i$ or $j$
(we choose one with equal probability) to a new node randomly chosen from the $N-2$
remaining nodes of the network.  This process is repeated for all the links present in the
original graph. For each value of $p$ we generate a sample of rewired networks over which averages are performed. In this way, we have new networks with a fraction $p$ of links different
from those of the original RGG. Besides, we only consider those resulting networks that
after the rewiring process have one single connected component.

The rewiring process is aimed to gradually (as $p$ varies from $0$ to $1$) destroy the
correlations between nodes present in the RGG so that in the limit $p=1$ we
obtain ER networks. We have constructed ensembles of $10^{3}$ rewired networks for
different values of $p$ and computed the average clustering coefficient and average path
lenght in these ensembles. In Figure \ref{fig1} we plot both quantities as a function of the
rewiring parameter, $\langle C\rangle (p)$ and $\langle L\rangle(p)$, normalized to the
corresponding values found in the RGG. The result obtained is a small-world transition as $p$ grows
from $0$ to $1$: while the average path length decreases very fast with the rewiring of a few links (this rewiring creates shortcuts), the clustering coefficent $\langle C\rangle (p)$ shows a slower decrease
and remain roughly the same as in the RGG until $p\simeq 10^{-2}$. Notice, however, that given the large value of $\langle C\rangle$ for the RGG, even for larger values of $p$, the clustering coefficient is very high.

\section{Results and Discussions}

To inspect how the synchronization transition of the $N$ Kuramoto
oscillators depends on the underlying topology, we have performed
extensive numerical simulations of the model in the RGG and the rewired
networks. Starting from $\lambda=0$, we increase it at small intervals.
The natural frequencies and the initial values of $\theta_i$ are
randomly drawn from a uniform distribution in the interval $(-1/2,1/2)$
and $(-\pi,\pi)$, respectively. Then, we integrate the equations of
motion Eq.\ (\ref{ks1}) using a $4^{th}$ order Runge-Kutta method over
a sufficiently large period of time to ensure that the order
parameter $r$ reaches a stationary value. The procedure is repeated
gradually increasing $\lambda$.

\begin{figure}[t]
\begin{center}
\epsfig{file=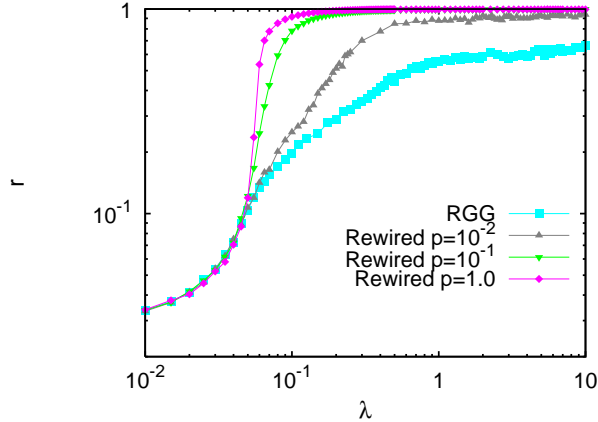,width=2.3in,angle=-90}
\end{center}
\caption{(color online) Order parameter $r$ as a function of the coupling strength $\lambda$.
    The global coherence is plotted for the original RGG and other three
    rewired networks corresponding to $p=10^{-2}$, $10^{-1}$ and $1$.}
\label{fig2}
\end{figure}

Figure \ref{fig2} shows the synchronization diagram  $r(\lambda)$ for the RGG and
three rewired versions ($p=10^{-2}$, $p=10^{-1}$ and $p=1$). For low values of
the coupling $\lambda$ it is observed that the RGG achieves local coherence ($r>0$)
slightly faster than its rewired versions. However, the transition to global
synchronization ($r = 1$) shows two clearly different behaviors. For both the RGG
and the randomized network with $p=10^{-2}$ the transition are extremely slow and
the global coherence is finally attained (not shown in figure) at very high values
of $\lambda$. This resistance of the network to achieve global synchronization
disappears for the other two rewired networks studied, $p=10^{-1}$ and $p=1$ (the
ER network). These two qualitative different behaviors in the synchronization transition
can be understood by looking at the evolution of the clustering coefficent in Fig.
\ref{fig1}. It seems clear that the high clustering coefficient present both in the
RGG and those networks with $p=10^{-2}$ is at the root of the extremely slow convergence
to $r=1$.

\begin{figure}[!htb]
\begin{center}
\epsfig{file=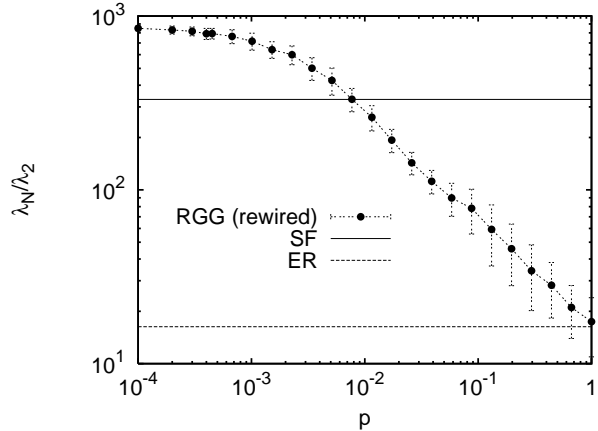,width=2.3in,angle=-90}
\end{center}
\caption{Synchronizability of the rewired networks as a function of the
    fraction of rewired links $p$. The synchronizability is measured as the
    eigenratio $\lambda_{N}/\lambda_{2}$ (the ratio between the largest and
    the first non-vanishing eigenvalues of the Laplacian matrix). We have also
    plotted the value of this eigenratio for SF and ER networks with the same
    number of nodes $N$ and average connectivity $\langle k\rangle$.}
\label{fig3}
\end{figure}

\begin{figure}[!htb]
\begin{center}
\epsfig{file=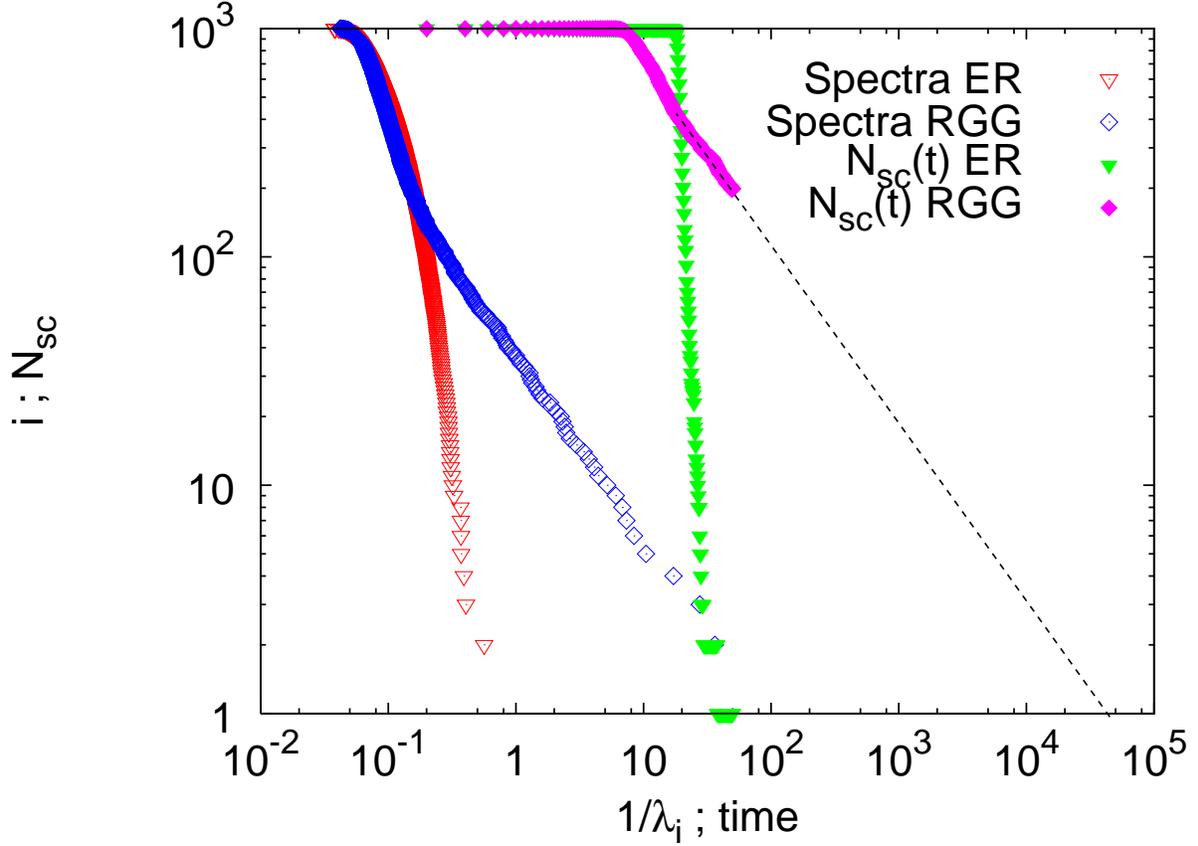,width=\columnwidth}
\end{center}
\caption{(color online) Index of the eigenvalue as a function of its numerical value for the RGG and for the
completely rewired one (leftmost curves) and number of synchronized components as a function of time for the
same networks (rightmost curves).}
\label{fig4}
\end{figure}


We next focus our attention on the stability of the synchronized state. As noted before, we can now abandon the specifics of a model and work for generic dynamical systems under just a few restrictions (for more 
details of the formalism and the conditions for its application, see, for instance 
[Boccaletti et al, 2006]
and references therein).
In a series of previous works 
[Barahona \& Pecora, 2002; Nishikawa et al., 2003; Motter et al., 2005; Donetti et al., 2005; Zhou et al., 2006;
Chavez et al., 2005],
it has been established a relation between the dynamical behavior of the units close to the synchronization manifold and the eigenvalues of the
(static) matrix that accounts for the individual dynamics.
As was first shown by Barahona and Pecora [2002],
the synchronizability (linear stability of the synchronized state)
depends on the ratio between the largest and the smallest non-zero eigenvalues of the Laplacian matrix, that is related
to the adjacency matrix by means of
\be
L_{ij}=k_i\delta_{ij}-A_{ij}.
\ee
Later on, this framework has been extended to a more general situation in which networks can be directed
and/or weighted and the Laplacian is no longer the matrix that appears
explicitly in the equations of motion. In any case there is a clear mapping between the ratio of the two
extremal eigenvalues and the stability of the synchronized state.

In a different scenario, Arenas et al. [2006]
showed that the relation between spectral properties
and dynamics goes far beyond the stability of the synchronized state.
In the case of Kuramoto oscillators for undirected and unweighted networks, as described in this paper,
the linearized dynamics
\begin{equation}
\frac{d\theta_i}{dt}=\omega_i + \sum_{j}
\sigma A_{ij}(\theta_j-\theta_i)= \omega_i + L_{ij}\theta_j \hspace{0.5cm} i=1,...,N
 \label{ks}
\end{equation}
is described exclusively in terms of the Laplacian matrix, for {\it identical} oscillators running with the same
intrinsic frequency, i.e., $w_i=w$ ($w$ can in turn be set to zero without loss of generality). Ordering the eigenvalues of the Laplacian
\be
0=\lambda_1\le \lambda_2 \le  \ldots \le \lambda_N,
\ee
they show that there is a striking similarity between the spectral representation (plotting the index of the eigenvalue as a function of the inverse of the numerical value) and the dynamical evolution (plotting the number of
synchronized components as a function of time) for networks of Kuramoto oscillators with a clear community structure.
In that case a plateau in the spectral representation (degeneracy of the eigenvalues) is
related to the synchronization stability of clusters of oscillators that form the topological communities.
Here it is important to notice that the approximation of phase oscillators coupled through the sine of the
phase difference is very close to its linear counterpart, the main difference being in the
initial stage of the evolution when phase differences are not small. But when clusters of synchronized oscillators form
the difference is so small in all pairs of interacting oscillators that the approximation is very good and
explains the accuracy of the mapping between the two representations 
[Arenas et al., 2006].

Another interesting finding is that the time needed for the whole system to synchronize can be shown to depend
mainly on the smallest nonzero eigenvalue $\lambda_2$. In the above picture it corresponds to the behavior at
long times. In 
[Almendral \& D\'{\i}az-Guilera, 2007]
it is analytically proven that this dependence for the linear
case is $T_{\mbox{\scriptsize{sync}}}\sim 1/\lambda_2$. For other dynamics, as the Kuramoto oscillators,
it is shown that this dependence is maintained although the prefactors can be relevant and we can have a
rough idea on the scaling of the time for a set of comparable networks but we cannot infer the time a single network
will need to get completely synchronized.

However, these two facts put together can be very helpful in understanding the dynamics of Kuramoto oscillators
in a network like RGG. In Fig. \ref{fig3} we plot the synchronizability for the original RGG and
the rewired versions. There we see that the ratio decays about two orders of magnitude from the original RGG
to the completely rewired, i.e. ER equivalent. Since the largest eigenvalue depends mainly on the
maximum degree of the nodes, we can infer that it is the second eigenvalue the responsible for such a decay.
This also means that the RGG network should need about two orders of magnitude more in time to get synchronized.
For the size of the networks we are considering and the accuracy necessary for any detailed discussion this
makes this time scale to be out of our available computer resources at this moment.
But we can make some predictions about this time, although this assumption has to be taken with care.
In Fig. \ref{fig4} we have plotted, as suggested above for networks with community structure, the index of the eigenvalue as a function of the inverse of its numerical value for the original RGG and for the completely rewired one.
We can see in these curves the different organization of the set of eigenvalues.
While for the ER graph there is a rapid decay, for the RGG there is a slow and systematic drop.
We can proceed in identifying the spectrum of the Laplacian with the synchronization dynamics, even  if there
is no community structure. This is what is represented in the other curves of Fig. \ref{fig4}, the number of
synchronized components as a function of time for the same two networks. This number is obtained following the method introduced in  
[Arenas et al., 2006].
Notice that the ER graph follows
a similar rapid drop, meaning that synchronization takes place in a very fast scale once the
initial readjustments of phase have taken place. But for the RGG we cannot proceed to complete synchronization because
of computer time limitations but we can notice the similarity in the slope of the two curves, clusters of oscillators
get synchronized following the spectrum of the Laplacian matrix. If we extrapolate this behavior we can infer
an approximation for the time needed for the RGG network to get completely synchronized.

Going back to the results depicted in Fig. \ref{fig3}, we see that the completely synchronized state in RGGs is the less stable of all the networks considered. Namely, the RGG performs worse than scale-free networks and random graphs in the region of the phase diagram where the coupling strength is high enough as to achieve full synchronization in all graphs. Therefore, if the synchronizability of a network is measured as the robustness of the completely synchronized state of the system, RGGs are not the best architecture. This behavior, however, can be radically changed by adding a few long range connections to the network. Through this rewiring mechanism, the synchronizability of RGG is improved well beyond that of scale-free networks as shown in Fig. \ref{fig3} for $p\ge 10^{-2}$.

\section{Conclusions}

We have explored the synchronization properties of RGGs. In order to inspect the onset of synchronization and partially synchronized states, we have made use of the Kuramoto model and showed that RGGs start to show some degree of synchrony roughly at the same critical value of the coupling strength found for random graphs. However, the former networks are harder to synchronize completely. Once the fully synchronized state is achieved, this state is the less robust compared to other classes of complex networks. A rewiring mechanism can easily solve the problem by adding just a few shortcuts to the network, which provokes both a faster landing on the fully synchronized state and an increase of its robustness.

In summary, the results here obtained allow to rank RGG as not particularly suitable for synchronization phenomena $-$ all the elements of the system are hard to synchronize and once this state is attained, it is the less robust. On one hand, these are bad news given the crucial role that synchronization processes plays in these systems. On the other hand, the fact that by simply adding some shortcuts to the system its synchronization performance greatly improves constitutes the good news. This is in particular the case in the context of wireless ad-hoc networks where such shortcuts can be achieved using advanced technologies such as transmit power control and directional antennas. Future research should confirm our preliminary insights by incorporating more details of the structure and dynamics of wireless systems and should also test whether or not the proposed rewiring mechanism is effective.

\section{acknowledgments}
  We thank A. Arenas for very helpful comments and
  discussions on the subject. Y.M. is supported by MEC through the Ram\'{o}n y Cajal Program. This work has been partially
  supported by the Spanish DGICYT Projects FIS-2006-13321, FIS2006-12781-C02-01, and
FIS2005-00337, and by the European NEST Pathfinder project GABA under contract 043309.


\section{References}

\noindent

Acebr\'{o}n, J. A., Bonilla, L. L., P\'{e}rez Vicente, C. J., Ritort, F., \& Spigler, R. [2005] 
``The Kuramoto model: a simple paradigm for synchronization phenomena'', 
\emph{Rev. Mod. Phys.} {\bf 77},137-185.\\

Almendral, A. \& {D\'{\i}az-Guilera}, A. [2007]
"Dynamical and spectral properties of complex networks,"
\emph{New J. Phys.} {\bf 9}, 187.

Arenas, A., D\'{\i}az-Guilera, A. \& P\'{e}rez-Vicente, C. J. [2006]
 ``Synchronization reveals topological scales in complex networks'',
\emph{ Phys. Rev. Lett.} {\bf 96}, 114102.\\

Barahona, M. \& Pecora, L.M [2002] ``Synchronization in Small-World
Systems'', \emph{Phys. Rev. Lett.} {\bf 89}, 054101.\\

Boccaletti, S., Latora, V., Moreno, Y., Chavez, M., \&
Hwang, D.-U. [2006] ``Complex networks: Structure and dynamics'',
\emph{Phys. Rep.} {\bf 424}, 175-308.\\

Bornholdt S. \& Schuster, H. G. [2003] (Editors) 
``Handbook of Graphs and Networks'' 
(Wiley-VCH, Germany).\\

Chavez, M., Hwang, D.-U., Amann, A., Hentschel, H. G. E. \&
Boccaletti, S. [2005] ``Synchronization is Enhanced in Weighted
Complex Networks'', \emph{Phys. Rev. Lett.} {\bf 94}, 218701.\\

Dall, J. \& Christensen, M. [2002]
"Random geometric graphs,"
\emph{Phys. Rev. E} {\bf 66}, 016121.

Donetti, L., P. I. Hurtado, P. I. \& Mu\~{n}oz, M. A. [2005]
``Entangled Networks, Synchronization, and Optimal Network Topology'',
\emph{Phys. Rev. Lett.} {\bf95}, 188701.\\

Dorogovtsev S. N. \& Mendes, J. F. F. [2003] 
``Evolution of Networks. From Biological Nets to the Internet and the WWW'' 
(Oxford University Press, Oxford, U.K.).\\

Glauche, I., Krause, W., Sollacher, R. \& Geiner, M. [2004]
"Distributive routing and congestion control in wireless multihop ad hoc communication networks,"
\emph{Physica A} {\bf 341}, 677.

G\'omez-Garde\~{n}es, J., Moreno, Y. \& Arenas, A. [2006a],
"{Paths to Synchronization on Complex Networks}",
\emph{Phys. Rev. Lett.}, {\bf 98}, {034101}.

G\'omez-Garde\~{n}es, J., Moreno, Y. \& Arenas, A. [2006b],
"{Synchronizability determined by coupling strengths and topology on Complex Networks}",
\emph{Phys. Rev. E}, {\bf 75}, {066106}

Hekmat, R. [2006]
{\em Adhoc Networks: Fundamental Properties and Network Topologies} (Springer-Verlag, Berlin)

Krause, W., Scholz, J. \& Geiner, M. [2006]
"Optimized network structure and routing metric in wireless multihop ad hoc communication,"
\emph{Physica A} {\bf 361}, 707.

Kuramoto, Y. [1984] ``Chemical oscillations, waves, and
turbulence'' (Springer-Verlag, New York).\\

Manrubia, S. C, Mikhailov, A. S., \& Zanette, D. H. [2004]
``Emergence of Dynamical Order. Synchronization Phenomena in
Complex Systems'' 
(World Scientific, Singapore).\\

Motter, A. E., Zhou, C. \& Kurths, J. [2005] ``Network
synchronization, diffusion, and the paradox of heterogeneity'',
\emph{Phys. Rev. E} {\bf 71}, 016116.\\

McGraw, P. N. \& Menzinger, M. [2005] ``Clustering and the
synchronization of oscillator networks'', 
\emph{Phys. Rev. E} {\bf 72},
015101(R).\\

Moreno, Y. \& Pacheco, A. F. [2004] 
``Synchronization of Kuramoto oscillators in scale-free networks'', 
\emph{Europhys. Lett.} {\bf 68}, 603-609.\\

Nekovee, M. [2007]
"Worm epidemics in wireless ad hoc networks,"
\emph{New J. Phys.} {\bf 9}, 189.

Nishikawa, T., Motter, A. E., Lai, Y.-C. \& Hoppensteadt, F. C. [2003]
``Heterogeneity in Oscillator Networks: Are Smaller Worlds Easier to Synchronize?'', 
\emph{Phys. Rev. Lett.} {\bf 91}, 014101.\\

Oh, E., Rho, K., Hong, H., \& Kahng, B. [2005] 
``Modular synchronization in complex networks'', 
\emph{Phys. Rev. E} {\bf 72}, 047101.\\

Penrose, M. [2003]
\emph{Random Geometric Graphs}  
(Oxford University Press, Oxford).

Strogatz, S. H. [2003] 
``Sync: The Emerging Science of Spontaneous Order'', 
(New York, Hyperion).\\

Winfree, A. T. [1990] 
``The geometry of biological time'',
(Springer-Verlag, New York)\\

Zhou, C., Motter, A. E. \& Kurths, J. [2006] 
``Universality in the Synchronization of Weighted Random Networks'', 
\emph{Phys. Rev. Lett.} {\bf 96}, 034101.\\

\end{document}